\documentclass[a4paper,11pt]{article}

\usepackage{fullpage}
\usepackage{times}
\usepackage{soul}
\usepackage{url}
\usepackage[utf8]{inputenc}
\usepackage[small]{caption}
\usepackage{graphicx}
\usepackage{xcolor}
\usepackage{amsmath}
\usepackage{booktabs}
\usepackage{algorithm}
\usepackage{enumitem}
\usepackage{todonotes}

\usepackage{amsthm}
\usepackage{amssymb}
\usepackage{algpseudocode}
\usepackage{cleveref}

\usepackage{bbm}

\newtheorem{theorem}{Theorem}

\newtheorem{lemma}[theorem]{Lemma}

\theoremstyle{definition}

\usepackage{natbib}

\newcommand{\bw}{\mathbf{w}}
\newcommand{\bx}{\mathbf{x}}
\newcommand{\by}{\mathbf{y}}
\newcommand{\bz}{\mathbf{z}}

\allowdisplaybreaks

\usepackage{authblk}

\title{\bf Characterizing Maximum Nash Welfare:\\ Envy-Freeness and Proportionality}

\author[1]{Karen Frilya Celine}
\author[1]{Warut Suksompong}
\author[2]{Nicholas Teh}

\affil[1]{National University of Singapore, Singapore}
\affil[2]{University of Oxford, UK}

\date{\vspace{-10mm}}

\begin{document}

\maketitle

\begin{abstract}
When allocating indivisible goods, the maximum Nash welfare rule is known to be the unique welfarist rule satisfying envy-freeness up to one good.
We strengthen this characterization by showing that envy-freeness up to one good can be replaced by substantially weaker fairness requirements: envy-freeness up to $k$ goods or proportionality up to $k$ goods, for any fixed $k$.
Our characterization holds even for profiles with identical goods. 
\end{abstract}

\section{Introduction} \label{sec:intro}

Fair division refers to the study of how scarce resources should be allocated among agents with potentially different preferences \citep{BramsTa96,Moulin03}.
A central challenge arises when the resources are \emph{indivisible goods} which cannot be split fractionally, such as course seats, housing units, or offices.
The indivisible nature of such goods renders it impossible to guarantee the classic fairness notions of \emph{envy-freeness} (no agent should envy another agent) and \emph{proportionality} (each agent should receive her proportionally fair share).
For example, if there are more agents than goods and every agent values every good positively, then any allocation leaves some agent empty-handed, and such an agent is necessarily envious and deprived of her proportional share.

Among the numerous methods that have been proposed for allocating indivisible goods, one of the most prominent is the \emph{maximum Nash welfare (MNW)} rule.
From each profile, MNW selects an allocation that maximizes the product of the agents' utilities, or equivalently, the sum of their logarithms.
\citet{CaragiannisKuMo19} famously demonstrated the ``unreasonable fairness'' of MNW by showing that whenever the utilities are additive, every allocation chosen by the rule satisfies \emph{envy-freeness up to one good (EF1)}: if an agent envies another agent, then there exists a good in the latter agent's bundle whose removal eliminates the former agent's envy.\footnote{The work of \citet{CaragiannisKuMo19} received the Kalai Prize in 2024 and has garnered over 1,000 citations according to Google Scholar.}
\citet{YuenSu23} further strengthened the case for MNW by characterizing it as the unique \emph{welfarist rule} that ensures EF1---a~welfarist rule selects an allocation maximizing some function of the agents' utilities.
This also extended an earlier characterization by \citet{Suksompong23} for \emph{additive} welfarist rules, which return an allocation maximizing the \emph{sum} of some function of the agents' utilities.\footnote{
Recently, \citet{BrandlSuTe26} characterized MNW under binary utilities (for such utilities, MNW is equivalent to any additive welfarist rule with a strictly concave function) using EF1 along with other properties.
\citet{CelineSuYu26} characterized additive welfarist rules satisfying EF1 for specific classes of profiles.
\citet{EckartPsVe26} investigated ``$p$-mean rules'' (MNW corresponds to the case $p = 0$) for profiles where the utilities are normalized.}

Despite the characterization of \citet{YuenSu23}, it remains conceivable that other welfarist rules achieve slightly weaker fairness guarantees than MNW.
For instance, \citet{MontanariScSu25} proved that the \emph{maximum harmonic welfare (MHW)} rule, which selects an allocation that maximizes the sum of harmonic utilities, ensures EF1 for profiles with integer utilities.
In fact, \citet{CelineSuYu26} illustrated that MHW can have certain advantages over MNW in terms of utilitarian welfare (i.e., the sum of agents' utilities).\footnote{See their Example~1.1 and the discussion afterward.}
A natural question is therefore whether there exist welfarist rules besides MNW that attain relaxations of EF1 such as \emph{envy-freeness up to $k$ goods (EF$k$)} or \emph{proportionality up to $k$ goods (PROP$k$)} for some fixed $k \ge 1$, perhaps even for specific numbers of agents.
We defer the definitions of these notions to \Cref{sec:prelim}, but remark here that PROP$k$ is fundamentally weaker than EF$k$.

In this note, we provide a complete negative answer to the question above.
Specifically, for every number of agents $n \ge 2$ and every integer $k \ge 1$, we show that MNW remains the unique welfarist rule that guarantees EF$k$ for all profiles with $n$ agents; the same characterization also holds for PROP$k$.
This means that MNW is, in a certain sense, significantly fairer than all other welfarist rules.
Moreover, our characterization holds even for profiles in which all goods are identical.\footnote{The construction used by \citet{YuenSu23} for EF1 (as well as the earlier one by \citet{Suksompong23} for additive welfarist rules) inherently relies on non-identical goods and cannot be extended to EF$k$ or PROP$k$ for $k > 1$.}

\section{Preliminaries} \label{sec:prelim}

Let $N = \{1, \ldots, n\}$ and $G = \{g_1, \ldots, g_m\}$ be the set of agents and the set of goods, respectively.
Each agent $i \in N$ has a \emph{utility function} $u_i : 2^G \to \mathbb{R}_{\geq 0}$; for a single good $g \in G$, we write $u_i(g)$ instead of $u_i (\{g\})$ for simplicity. 
We assume that the utility functions are additive, that is, $u_i (G') = \sum_{g \in G'} u_i (g)$ for all $i \in N$ and $G' \subseteq G$. 
A \emph{profile} consists of $N$, $G$, and $(u_i)_{i \in N}$. 
A profile is said to be \emph{identical-good} if for all $i\in N$ and $g,g'\in G$, it holds that $u_i(g) = u_i(g')$.

An \emph{allocation} $A = (A_1, \ldots, A_n)$ is an ordered partition of $G$ into $n$ (possibly empty) bundles such that for each $i\in N$, bundle $A_i$ is allocated to agent $i$.
An agent $i\in N$ receives utility $u_i(A_i)$ from allocation~$A$.
A~profile is \emph{positive-admitting} if there exists an allocation such that every agent receives positive utility.
For a positive integer $k$, an allocation~$A$ is said to be
\begin{itemize}
\item \emph{envy-free up to $k$ goods (EF$k$)} if for every pair of agents $i,j\in N$, there exists $B\subseteq A_j$ with $|B| \le k$ such that $u_i(A_i) \ge u_i(A_j\setminus B)$;
\item \emph{proportional up to $k$ goods (PROP$k$)} if for every agent $i\in N$, there exists $B\subseteq G\setminus A_i$ with $|B|\le k$ such that $u_i(A_i\cup B) \ge u_i(G)/n$.
\end{itemize}
For any positive integer $k$,  EF$k$ implies\footnote{See the proof of Proposition~2.2(b) of \citet{ManurangsiSu22}.} PROP$k$.
A \emph{rule} is a function that maps any given profile to an allocation. 

Given $n \geq 2$, a \emph{welfare function} is a non-decreasing function\footnote{Note that $-\infty$ is necessary in the codomain.
For example, for the representation of MNW as maximizing the sum of the logarithms of the agents' utilities, if some agent has zero utility, the value of the function is $-\infty$.} $f_n : [0, \infty)^n \to [-\infty, \infty)$.
The \emph{welfarist rule with (welfare) function $f_n$} chooses, from each profile with $n$ agents, an allocation $A$ that maximizes the \emph{welfare} $f_n(u_1 (A_1), \ldots, u_n (A_n))$; if there are multiple such allocations, the rule may choose one arbitrarily.

\section{The Result} \label{sec:result}

Before proceeding to our characterization, we first prove an auxiliary lemma.

\begin{lemma}
\label{lem}
Fix a positive integer $n \ge 2$.
Let $f_n : [0, \infty)^n \to [-\infty, \infty)$ be a non-decreasing function that is continuous on $(0, \infty)^n$.
Suppose that for all $\bx = (x_1,\dots,x_n) \in (0,\infty)^n$ and $\by = (y_1,\dots,y_n) \in (0,\infty)^n$ such that $$\sum_{i\in N} \frac{y_i}{x_i} = n,$$
it holds that $f_n(\bx) \ge f_n(\by)$.
Then, for all $\bx = (x_1,\dots,x_n) \in (0,\infty)^n$ and $T > 0$, we have
\begin{align}
\label{eq:multiply}
f_n(x_1,x_2,x_3,\dots,x_n) = f_n\left(Tx_1,\frac{x_2}{T},x_3,\dots,x_n\right).
\end{align}
More generally, if we multiply one entry of $\bx = (x_1,\dots,x_n)$ by an arbitrary $T > 0$ and divide another entry by the same~$T$, the value of~$f_n$ does not change.
\end{lemma}

\begin{proof}
Let $\bx = (x_1,\dots,x_n)$ and $T > 0$, and let $\widehat{\bx} = (Tx_1, x_2/T, x_3, \dots, x_n)$ and $s = \ln T \in (-\infty, \infty)$.
For each integer $S > |s|$ and each $t\in\{0,1,\dots,S\}$, define
\[
\bx^{S,t} = \left( \left(1+\frac{s}{S}\right)^t x_1, \left(1-\frac{s}{S}\right)^t x_2, x_3, \dots, x_n  \right).
\]
Note that all entries of $\bx^{S,t}$ are positive.
Also, for each $t\in\{1,\dots,S\}$, 
\[
\sum_{i\in N} \frac{x^{S,t}_i}{x^{S,t-1}_i} 
= \left(1 + \frac{s}{S}\right) + \left(1 - \frac{s}{S}\right) + (n-2) = n,
\]
so the assumption of the lemma implies that $f_n(\bx^{S,t-1}) \ge f_n(\bx^{S,t})$.
Applying this for all $t\in\{1,\dots,S\}$, we get $f_n(\bx^{S,0}) \ge f_n(\bx^{S,S})$, that is,
\[
f_n(\bx) \ge f_n\left( \left(1+\frac{s}{S}\right)^S x_1, \left(1-\frac{s}{S}\right)^S x_2, x_3, \dots, x_n  \right).
\]
Recall that $\lim_{S\rightarrow\infty} \left(1+\frac{s}{S}\right)^S = e^s$ and $\lim_{S\rightarrow\infty} \left(1-\frac{s}{S}\right)^S = e^{-s}$.
Taking the limit $S\rightarrow\infty$ on both sides of the inequality above and using the continuity of~$f_n$, we get
\[
f_n(\bx) \ge f_n(e^s x_1, e^{-s} x_2, x_3, \dots, x_n) = f_n\left(Tx_1, \frac{x_2}{T}, x_3, \dots, x_n\right) = f_n(\widehat{\bx}).
\]
Similarly, by starting with $\widehat{\bx}$ instead of $\bx$ and repeating the same arguments with $-s$ instead of $s$, we obtain the reverse inequality $f_n(\widehat{\bx}) \ge f_n(\bx)$.
Combining the two inequalities yields \eqref{eq:multiply}.

The same reasoning also implies that if we start with $\bx = (x_1,\dots,x_n)$, multiply one entry by an arbitrary $T > 0$, and divide another entry by the same~$T$, the value of~$f_n$ does not change.    
\end{proof}

We state our characterization in the following theorem.

\begin{theorem} 
\label{thm:main}
Fix positive integers $n \geq 2$ and $k \ge 1$. 
Let $f_n$ be a welfare function that is continuous and strictly increasing\footnote{As observed by \citet[Footnote~3]{YuenSu23}, in order for the characterization to hold, one needs to assume that $f_n$ is strictly increasing on $(0, \infty)^n$, in addition to being non-decreasing on $[0,\infty)^n$.} on $(0, \infty)^n$. 
Then, the following three statements are equivalent:
\begin{enumerate}[label=(\alph*)]
    \item For every positive-admitting profile with $n$ agents, every allocation that can be chosen by the welfarist rule with function $f_n$ is EF$k$.
    \item For every positive-admitting identical-good profile with $n$ agents, there exists an EF$k$ allocation that can be chosen by the welfarist rule with function~$f_n$.
    \item The following two statements hold for $f_n$:
    \begin{enumerate}[label=(\roman*)]
        \item There exists a strictly increasing and continuous function $q : (0, \infty) \to (-\infty, \infty)$ such that $f_n(x_1, x_2, \ldots, x_n) = q(x_1 x_2 \cdots x_n)$ for all $x_1, \ldots, x_n > 0$.
        \item The inequality $f_n(x_1, x_2, \ldots, x_n) > f_n(y_1, y_2, \ldots, y_n)$ holds for all $x_1, \ldots, x_n > 0$ and $y_1, \ldots, y_n \geq 0$ satisfying $\prod_{i\in N} y_i = 0$.
    \end{enumerate}
\end{enumerate}
The same equivalence holds when EF$k$ is replaced by PROP$k$ in both (a) and (b).
\end{theorem}

\begin{proof}
Fix $n$ and $k$ as in the theorem statement.
The implication (a) $\Rightarrow$ (b) is trivial for both EF$k$ and PROP$k$.
For the implication (c) $\Rightarrow$ (a), suppose that $f_n$ satisfies~(c).
Then, given a positive-admitting profile with $n$~agents, every allocation that can be chosen by the welfarist rule with function~$f_n$ can also be chosen by MNW.
Such an allocation must satisfy EF1 \citep{CaragiannisKuMo19}, and therefore EF$k$ and PROP$k$ for every $k\ge 1$.

It remains to establish the implication (b) $\Rightarrow$ (c).
Since EF$k$ implies PROP$k$, it suffices to prove this implication for PROP$k$.
Assume that $f_n$ satisfies (b) with PROP$k$ instead of EF$k$.
We will show that both statements (i) and (ii) of (c) hold.

Fix an arbitrary $\bx = (x_1,\dots,x_n) \in (0,\infty)^n$.
For each integer $L > k$, consider an identical-good profile with $|G| = nL$ goods such that $u_i(g) = x_i/L$ for all $i\in N$ and $g\in G$.
Clearly, this profile is positive-admitting.
Since $f_n$ satisfies~(b), there exists a PROP$k$ allocation~$A^L$ that can be chosen by the welfarist rule with function~$f_n$.
In particular, since the profile is identical-good, the definition of PROP$k$ implies that each agent must receive at least $L-k$ goods.
As there are $nL$ goods in total, this also means that each agent can receive at most $L+(n-1)k$ goods.
Hence, for each $i\in N$, it holds that $$\frac{L-k}{L}\cdot x_i\le u_i(A_i^L) \le \frac{L+(n-1)k}{L}\cdot x_i,$$
and so $\lim_{L\rightarrow\infty} u_i(A_i^L) = x_i$.
Letting $\bz^L = (u_1(A_1^L),\dots,u_n(A_n^L))$, we have $\lim_{L\rightarrow\infty} \bz^L = \bx$.

Next, consider any fixed $\by = (y_1,\dots,y_n) \in (0,\infty)^n$ satisfying $\sum_{i\in N} y_i/x_i = n$.
Let $d_i^L = \lfloor Ly_i/x_i\rfloor$ for each $i\in N\setminus\{n\}$, and let $d_n^L = nL - \sum_{i\in N\setminus\{n\}} d_i^L$.
For each $i\in N\setminus\{n\}$, since 
\[
\frac{y_i}{x_i} 
\ge \frac{d^L_i}{L}
> \frac{Ly_i/x_i - 1}{L}
= \frac{y_i}{x_i} - \frac{1}{L},
\]
we have $\lim_{L\rightarrow\infty} d^L_i/L = y_i/x_i$.
Moreover, since $\sum_{i\in N}d^L_i = nL$, it holds that 
\[
\lim_{L\rightarrow\infty} \frac{d_n^L}{L} 
= \lim_{L\rightarrow\infty} \frac{nL - \sum_{i\in N\setminus\{n\}} d_i^L}{L} 
= n - \sum_{i\in N\setminus\{n\}}\lim_{L\rightarrow\infty}\frac{d_i^L}{L} = n - \sum_{i\in N\setminus\{n\}}\frac{y_i}{x_i} = \frac{y_n}{x_n}.
\]
Thus, $\lim_{L\rightarrow\infty} d^L_i/L = y_i/x_i$ for every $i\in N$.

Consider an allocation $B^L$ where each agent $i\in N$ receives $d_i^L$ goods.
Observe that $\lim_{L\rightarrow\infty} u_i(B^L_i) = \lim_{L\rightarrow\infty} d^L_i\cdot x_i/L = y_i$ for each $i\in N$.
Letting $\bw^L = (u_1(B^L_1),\dots,u_n(B^L_n))$, we have $\lim_{L\rightarrow\infty} \bw^L = \by$.
Since the welfarist rule with function~$f_n$ can choose~$A^L$, it must hold that $f_n(\bz^L) \ge f_n(\bw^L)$ for each~$L$.
Taking the limit $L\rightarrow\infty$ and applying the continuity of~$f_n$, we get $f_n(\bx) \ge f_n(\by)$.
Hence, 
\begin{align*}
f_n(\bx) \ge f_n(\by) \text{ for every } \by = (y_1,\dots,y_n) \in (0,\infty)^n \text{ such that } \sum_{i\in N} \frac{y_i}{x_i} = n.
\end{align*}
By \Cref{lem}, if we start with any $\bx = (x_1,\dots,x_n)$, multiply one entry by an arbitrary $T > 0$, and divide another entry by the same~$T$, the value of~$f_n$ does not change.

We are now ready to prove statement~(i) of~(c).
Suppose that $\bx = (x_1,\dots,x_n) \in (0,\infty)^n$ and $\bx' = (x'_1,\dots,x'_n) \in (0,\infty)^n$ are such that $\prod_{i\in N}x_i = \prod_{i\in N}x'_i$.
Starting from~$\bx$, we can modify one entry at a time from left to right to match~$\bx'$, each time modifying the last entry to keep the product fixed.
Specifically, for each $i\in N\setminus\{n\}$, we multiply the $i$-th entry by $x'_i/x_i$ and divide the $n$-th entry by the same factor.
By the previous paragraph, the value of~$f_n$ does not change during this process.
Since the product of all $n$~entries remains the same, the equality of products implies that the $n$-th entry becomes $x'_n$ at the end of the process.
This means that $f_n(\bx) = f_n(\bx')$.
In particular, $f_n(x_1,x_2,\dots,x_n) = f_n(x_1x_2\cdots x_n,1,1,\dots,1)$.
Let $q : (0, \infty) \to (-\infty, \infty)$ be such that $q(t) = f_n(t,1,1,\dots,1)$ for each $t > 0$.
Since $f_n$ is strictly increasing and continuous, so is $q$.
Moreover, $q$ is real-valued because $q(t) > q(t/2) \ge -\infty$ for every $t > 0$.
For any $x_1,\dots,x_n > 0$, we have 
\[
f_n(x_1,x_2,\dots,x_n) = f_n(x_1x_2\cdots x_n,1,1,\dots,1) = q(x_1x_2\cdots x_n).
\]
This establishes~(i).

It remains to prove statement~(ii) of~(c).
Let $x_1,\dots,x_n > 0$ and $y_1,\dots,y_n \ge 0$ be such that $\prod_{i\in N} y_i = 0$.
Without loss of generality, assume that $y_1 = \dots = y_r = 0$ and $y_{r+1},\dots,y_n > 0$ for some $r\in N$.
Let $X = \prod_{i\in N}x_i > 0$ and $Y = \prod_{i=r+1}^n y_i > 0$, where the empty product is taken to be~$1$ if $r = n$.
Define $z_1,\dots,z_n$ by 
\[
z_i = 
\begin{cases}
\left(\frac{X}{2Y}\right)^{1/r} & \text{if } i \le r; \\
y_i & \text{if } i > r.
\end{cases}
\]
Note that $z_i \ge y_i$ and $z_i > 0$ for all $i\in N$, and $\prod_{i\in N}z_i = X/2$.
Since $f_n$ is non-decreasing on $[0,\infty)^n$ by definition of a welfare function, and admits a strictly increasing product representation~$q$ on positive entries as proven earlier, we get
\[
f_n(y_1,\dots,y_n)
\le f_n(z_1,\dots,z_n)
= q(X/2)
< q(X)
= f_n(x_1,\dots,x_n).
\]
This establishes (ii) and completes the proof of the theorem.
\end{proof}

Note that if $f_n$ satisfies~(c), then for any positive-admitting profile, an allocation can be chosen by the welfarist rule with function~$f_n$ if and only if it can be chosen by MNW.
Hence, any such rule must be effectively equivalent to MNW.\footnote{For profiles that are not positive-admitting, MNW requires additional tie-breaking to guarantee EF1 \citep{CaragiannisKuMo19}.}

Finally, we observe the following consequence of \Cref{thm:main}.
Apart from MNW, there cannot exist a welfarist rule for any number of agents defined by a family of functions $\{f_n\}_{n\ge 2}$, each satisfying the conditions of \Cref{thm:main}, together with a function $h : \mathbb{Z}_{\ge 2} \to \mathbb{Z}_{\ge 1}$, with the following property: for each number of agents~$n$, the allocation produced by the rule always satisfies envy-freeness (or proportionality) up to $h(n)$ goods.
Indeed, \Cref{thm:main} precludes this possibility even for any specific~$n$.

\subsection*{Acknowledgments}

This work was partially supported by the Ministry of Education, Singapore, under the Academic Research Fund Tier 1 (FY 2026) grant number 251RES2604, and by an NUS Start-up Grant.
We thank the anonymous reviewer for valuable comments.

\subsection*{Declaration of Generative AI Use}

ChatGPT 5.6 Pro was used to help explore proof ideas.
The authors wrote the entire article and take full responsibility for its content.

\bibliographystyle{plainnat}
\bibliography{bib}

\end{document}